\documentclass[prd,showpacs,amsmath,showkeys,twocolumn,floatfix]{revtex4} 
\usepackage{epsfig,dcolumn}
\usepackage{graphicx}
\usepackage{graphicx}
\usepackage{bm}

\def\qq{{q{\bar q}}}
\def\x{{\bf x}}
\def\y{{\bf y}}
\def\z{{\bf z}}
\def\k{{\bf k}}
\def\q{{\bf q}}
\def\p{{\bf p}}

\def\A{{\bf A}}
\def\B{{\bf B}}

\def\R{{\bf R}}

\def\lsim{\mathrel{\rlap{\lower4pt\hbox{\hskip1pt$\sim$}}
    \raise1pt\hbox{$<$}}}
\def\gsim{\mathrel{\rlap{\lower4pt\hbox{\hskip1pt$\sim$}}
    \raise1pt\hbox{$>$}}}
\begin{document}


\title{ Coulomb energy and gluon distribution in the presence of static sources. }

\author{ Adam P. Szczepaniak and Pawel Krupinski}
\affiliation{ Physics Department and Nuclear Theory Center \\
Indiana University, Bloomington, Indiana 47405 }

\date{\today}

\begin{abstract}
We compute the energy of the ground state and a low lying excitation of the gluonic field in the presence 
  of static quark -anti-quark ($\qq$)  sources. We show that for 
separation between the sources less then  a few fm the gluonic 
ground state of the static $\qq$ system can be well  
described in terms of a  mean field  wave functional with 
 the excited states corresponding to a single quasi-particle 
excitation of the gluon field.  We also discuss the  role of 
 many particle excitations relevant for large separation 
 between sources.
  \end{abstract}

\pacs{11.10Ef, 12.38.Aw, 12.38.Cy, 12.38.Lg}





\maketitle
\section{Introduction} 
Recent lattice simulations lead to many new theoretical insights into  the dynamics of low-energy gluon  modes~\cite{Juge:1997nc,Juge:2002br,Takahashi:2004rw,Luscher:2004ib,Cornwall:2004gi,Greensite:2001nx}. In the quenched approximation aspects of confinement emerge from  studies  of the gluonic spectrum produced by static color sources.  In the following we will focus on the pure gluon dynamics  
  (the role of dynamical quarks in the screening of confining gluonic strings has recently been studied in ~\cite{Bali:2005fu}).

Lattice studies indicate that with relative separations between two color sources, $R \gsim 1.7 \mbox{ fm}$ ,  the ground state energy obeys Casimir scaling~\cite{Bali:2000un,Juge:2004xr}. This means that the spectrum of gluon modes generated by static color sources depends on the dimension of the color representation of the sources  rather than on the N-ality of the representation (which is related to the transformation property of a representation with respect to the group center)~\cite{Greensite:2003bk}. For example, for two sources in the fundamental representation, lattice computations show, as expected, that energy grows linearly with the separation between the sources. However, also for sources in the adjoint representation (with N-ality of zero), lattice  produces a linearly rising potential, even though for vanishing N-ality screening is expected to saturate the potential.  Screening comes from the production of gluon pairs 
  (glueballs) which vanishes  in the limit of a large number of colors. Casimir scaling is thus telling us that there is, at least in the energy range relevant for hadronic phenomenology, a simple, universal (source independent) description of  the confining string.

The lattice spectrum of gluonic modes generated by sources in the fundamental representation, {\it i.e.}, a  static quark-antiquark ($\qq$) pair, has been extensively studied in ~\cite{Juge:1997nc,Juge:2002br}. The ground state energy, which as a function of the $\qq$ separation is well represented by the Cornell, "Coulomb+linear" potential and the spectrum of excited gluonic modes have been computed. The excited gluonic modes lead to excited adiabatic potentials between the sources in the sense  of the Born-Oppenheimer approximation with the quark sources and gluonic field corresponding to the slow and fast degrees of freedom, respectively~\cite{Juge:1999ie,Juge:1999aw}. The gluonic wave functional  of these modes can be classified analogously to that of a diatomic molecule. The good quantum numbers are: $\Lambda=0 (\Sigma) ,1 (\Pi) ,2 (\Delta) ,\cdots$  which give the total gluon spin projection along the $\qq$ axis, $PC=+1(g), -1(u)$ which correspond to the product of  gluon parity and charge conjugation, and $Y=\pm 1$ which describes parity under reflection in a plane containing the $\qq$ axis.  The ground state corresponds to $\Lambda^{Y}_{PC}=\Sigma^+_g$. The lattice calculations show that the first excited state has the $\Pi_u$ symmetry (for $\Lambda\ne 0$, $Y=\pm 1$ states are degenerate) and thus  has $PC= -1$.

The lattice spectrum of gluonic excitations is well reproduced by  the bag model~\cite{Hasenfratz:1980jv,Juge:1997nd} . The crucial feature of the model that makes this possible is the boundary condition, which requires  the longitudinal component of the chromo-electric  and transverse components of the chromo-magnetic field of the free gluon inside the cavity to  vanish at the boundary of the bag.  This results in the TE mode with pseudo-vector, $J^{P,C} = 1^{+,-}$, quantum numbers having the lowest energy, which leads to the  $\Pi_u$ adiabatic potential being the lightest from among the excited  gluonic states in the $\qq$ system. In another model, the non-relativistic flux tube model~\cite{Isgur:1984bm} , the $PC=+1$ quantum numbers of the low-lying gluon mode result from associating a negative parity and a positive charge conjugation to the lowest order transverse phonon (unlike that of a vector field). This also  results in the $\Pi_u$ quantum numbers for the first excited adiabatic potential.  Finally in a QCD based  quasi-particle picture the intrinsic quantum numbers of the quasi-gluons are, $J^{P,C} = 1^{-,-}$, that of a transverse vector field~\cite{Horn:1977rq,Swanson:1998kx} . If the first excited adiabatic potential between $\qq$ sources is associated with a single quasi-gluon excitation and this quasi-gluon interacts via normal two-body forces with the sources, then, one expects the quasi-gluon ground state wave function  to be in an orbital $S$-wave, which, in turn, leads to the net $PC=+1$ and the $\Pi_g$ symmetry for this state. This is in contradiction with the lattice data as noted in~\cite{Swanson:1998kx}. The bag model and the flux tube model give the right ordering of the spectrum of low lying gluonic excitations, even though they are based on very different microscopic representations of the gluonic degrees of freedom.

There are indications from lattice simulations of various gauge models  that the adiabatic potentials  approach that of the flux tube, or better string-like spectrum for $\qq$ separations larger then $R \gsim 3 \mbox{ fm}$~\cite{Juge:2003ge}, however, the situation for QCD is far less clear~\cite{Juge:2002br}. In particular, for large separations  between the sources the splitting  between nearby string excitations is expected to fall off as $\propto \pi/R$. The lattice results indicate, however,  that the spacing between the adiabatic potentials is close to constant. At distances $R \lsim 0.2 \mbox{ fm}$ the flux tube model becomes inadequate while QCD is expected to become applicable. For example as $R \to 0$, the Coulomb potential between   the quark and the anti-quark in the color octet is repulsive, and, indeed, the results of lattice calculations do seem to have that trend.  The bag model attempts to combine the perturbative and long range, collective dynamics by using a free fled theory inside a spherically symmetric bag and deforming the bag to a string like shape as the separation between the sources increases.  A self consistent  treatment of bag and gluon degrees of freedom is, however, lacking.

 Another model which aims at relating the string-like excitations at large $\qq$ separations with 
the QCD gluon degrees of freedom is the gluon chain model~\cite{Thorn:1979gu,Greensite:2001nx} and versions  thereof~\cite{Szczepaniak:1996tk}. The model is based on the assumption that as the separation between the sources increases pairs of constituent gluons are created to screen the charges in such a way that the Fock space  is dominated by a state with a number of constituent  gluons, which grows with the $\qq$ separation. Recently, support for the gluon chain model came from lattice studies of the Coulomb energy of the $\qq$ pair~\cite{Greensite:2004ke,Greensite:2003xf}. As  shown  in ~\cite{Zwanziger:2002sh},  at fixed $R$,  Coulomb energy bounds the true exact (from Wilson line) energy from above. The Coulomb energy is defined as the expectation value of the Coulomb potential in a state obtained by adding the $\qq$ pair to the exact ground state of the vacuum, {\it i.e.}, without taking into account vacuum polarization by the sources.  The addition of sources changes the vacuum wave functional by creating constituent gluons as described by the gluon chain model.

In this paper we discuss the structure of the $\qq$ state in terms of physical, transverse gluon degrees of freedom. In particular, we focus on the importance of constituent gluons in describing the excited adiabatic potentials. For simplicity and to make our arguments clearer,  we  concentrate on excited adiabatic potentials  of single, $\Lambda^Y_{PC} = \Sigma+_g$, symmetry. A description of the complete  spectrum of excited potentials will be presented in a following paper.  Our main finding here is that a description based on a single (few) constituent gluon excitation is valid up to $R \sim \mbox{few fm} $, with the gluon chain turning in, most likely, at asymptotically large $\qq$ separations. Consequently, we  show how the gluon chain model can  emerge in the basis of transverse gluon Fock space.

  In Section~II we review the Coulomb gauge formulation of QCD and introduce the Fock space of quasi-gluons. In Section~III we review the computation of the ground state  and the excited $\Sigma^+_g$ potentials. There we also discuss the role of multi-particle Fock sectors and a schematic model of the  gluon chain. A summary and outlook are given in Section~IV. 
  
  \section{Coulomb gauge QCD}

In the Coulomb gauge gluons have only physical degrees of freedom. For all color components $a=1,\cdots,N^2_C-1$ the gauge condition, $\bm{\nabla}\cdot{\bf A}^a(\x) = 0$,   eliminates the longitudinal  degrees of freedom and the scalar potential, $A^{0,a}$, becomes dependent on the transverse components through Gauss's law~\cite{Christ:1980ku}.  The canonical momenta, $\bm{\Pi}^a(\x)$, satisfy  $[\Pi^a_i(\x),A_j^b(\y)] = -i \delta_{ab}\delta^{ij}_T(\bm{\nabla})\delta^3(\x-\y)$ where $\delta^{ij}_T(\bm{\nabla}) = \delta^{ij} - \nabla^i\nabla^j/\bm{\nabla}^2$; in the Shr{\"o}dinger representation, the momenta  are given by $ \bm{\Pi}^a(\x) = -i\delta/\delta {\bf A}^a(\x)$. More discussion of the topological properties of the fundamental domain of the gauge variables can be found in ~\cite{vanBaal:1997gu}.    The full Yang-Mills (YM) Hamiltonian with gluons coupled to static $\qq$ sources  in the fundamental representation is given by,

 \begin{equation}
 H = H_0 + H_{Qg} + H_{QQ},
 \end{equation}
where $H_0$ is the YM Hamiltonian containing the kinetic term and interactions between transverse gluons. The explicit form of the YM Hamiltonian, $H_0$, can be found in ~\cite{Christ:1980ku}.  The coupling between $\qq$ sources and the transverse gluons, $H_{Qg}$, is explicitly  given by,

 \begin{equation}
 H_{Qg} = \int d\x d\y \rho_Q^a(\x) K[\A](\x,a;\y,b) \rho^b(\y), 
 \end{equation}
 where $\rho_Q =  h^{\dag}(\x)T^a h(\x) - \eta^{\dag}(\x)T^{*a} \eta(\x)$ is the color density of the sources with $h$ and $\eta$ representing the static quark and anti-quark annihilation operators, respectively; $\rho = -f_{abc} {\cal J}^{-1} \bm{\Pi}^b(\x) {\cal J} \cdot \A^c(\x)$ is the gluon charge density operator and $K$ is the non-abelian Coulomb kernel,  
 \begin{equation} 
 K[\A](\x,a;\y,b) = {{g^2}\over {4\pi}}  \int d\z {{  (1 - \lambda)^{-2}(\x,a;\z,b) } \over {|\z - \y|}} ,
  \end{equation}
with the matrix elements of $\lambda$ given by $(1-\lambda)(\x,a;\y,b) =\delta_{ab}\delta^3(\x-\y) - g f_{acb}  \bm{\nabla}_y (1/|\x - \y|) \A^c(\y)$. The Faddeev-Popov (FP) operator, $(1-\lambda)$, determines  the curvature of the gauge manifold specified by the FP determinant, ${\cal J} = \det(1-\lambda)$. 
 Finally, the interaction between the heavy sources, $H_{QQ}$,  is given by 
\begin{equation}
H_{QQ} = {1\over 2} \int d\x d\y \rho_Q^a(\x) K[\A](\x,a;\y,b) \rho^b_Q(\y).
\end{equation}
The Coulomb kernel  is a complicated function of the transverse gluon field. When $H_{Qg}$ and $H_{QQ}$ are  expanded in powers of the coupling constant, $g$, they lead to an infinite series of terms proportional to powers of $\A$. The FP determinant also introduces additional interactions. All  these interactions involving gluons in the Coulomb potential are responsible for binding constituent gluons to the quark sources. 

\subsection{ Fock space basis} 
The problem at hand is to find the spectrum of $H$ for a system containing a $\qq$ par,  
\begin{equation}
H |R,N \rangle = E_N(R) |R,N\rangle. \label{qq} 
\end{equation}
In the Shr{\"o}dinger representation, the eigenstates can be written as, 
\begin{equation}
|R,N\rangle = \int D[\A^a(x)] {\cal J}[A] \Psi^N_{ij}[\A^a(\x)] | {R\over 2}{\hat\z},i,-{R\over 2}\hat{\z},j;\A \rangle, 
\label{qqwf}
\end{equation}
with 
\begin{equation}
| {R\over 2} {\hat\z},i-{R\over 2} \hat{\z},j;\A \rangle = h^{\dag}_i({R\over 2} \hat{\z}) 
\eta^{\dag}_j(-{R\over 2}\hat\z) |\A \rangle
\end{equation}
describing a state containing a quark at position $R{\hat\z}/2$ and color $i$ and an anti-quark at position $-R\hat{\z}/2$ and color $j$. We keep quark spin degrees of freedom implicit since, for static quarks, the Hamiltonian is spin-independent. The eigenenergies, $E_N(R)$, correspond to  the adiabatic potentials discussed in Section~I with $N$ labeling  consecutive excitations and spin-parity, $\Lambda^Y_{PC}$, quantum numbers of the gluons in the static $\qq$ state. 

The vacuum without sources, denoted by $|0\rangle$,  in the Shr{\"o}dinger representation is given by,  

 \begin{equation}
 |0\rangle = \int D[\A^a(\x)]{\cal J}[A]  \Psi_0[\A^a(\x)] |\A \rangle , \label{0} 
 \end{equation}
  and satisfies  $ H_0 |0\rangle = E_{vac}|0\rangle$. 

The eigenenergies, $E_N(R)$,  in Eq.~(\ref{qq}) contain contributions from disconnected diagrams which sum up to the energy of the vacuum, $E_{vac}$. In the following, we will focus on the difference, $E_N(R) \to E_N(R) - E_{vac}$, and ignore disconnected contributions in the matrix elements of $H$. 

Instead of using  the Shr{\"o}dinger representation, it is convenient to introduce a Fock space for quais-particle-like gluons~\cite{Reinhardt:2004mm,Feuchter:2004mk,Szczepaniak:2003ve,Szczepaniak:2001rg}. These are defined in the standard way, as excitations built from a gaussian (harmonic oscillator) ground state.  Regardless of the choice of parameters of such a gaussian ground state, the set of all quasi-particle excitations forms a complete basis. 
 We will optimize this  basis by minimizing the  expectation value of the Hamiltonian in such a  gaussian ground state. We will then use this variational state to represent the physical vacuum and use it  in place of $|0\rangle$ and $\Psi_0[\A]$.  The unnormalized variational wave functional is given by,  $ \Psi_0[\A] = \langle \A| 0\rangle$, 
 \begin{equation}
 \Psi_0[\A] = \exp\left( -{1\over 2} \int {{d \k} \over {(2\pi)^3}} \A^a(\k) \omega(|\k|) \A^a(-\k) \right), 
 \label{psi0}
 \end{equation}
 where $\A^a(\k)  = \int d\x \exp(-i\k\cdot \x) \A^a(\x)$ and the  gap function, $\omega(|\k|)$ plays the role of the variation parameter. The computation of the expectation value of $H_0$ in $\Psi_0$ given above, 
   was  described in Ref.~\cite{Szczepaniak:2001rg}. In the following  we will summarize the main points.  
 
 The expectation value  of $\langle 0|H_0|0\rangle$ can be written in terms of 
   functional integrals over $D[\A^a(\x)]$ with the measure ${\cal J}[A]$. The functionals to be integrated   are products of $H_0 = H_0(\bm{\Pi},\A)$ and the wave functional $|\Psi_0[\A]|^2$. For example  the contribution to $\langle 0 |H_0 | 0 \rangle$ from the $g=0$ component of the transverse chromo-magnetic field density, $\langle B^2 \rangle =  \langle 0| \int d\x [\B^a(\x)]^2|0 \rangle/\langle 0|0\rangle$, is given by,  

 \begin{eqnarray}
 \langle B^2 \rangle & =  &
 \int D\A  {\cal J}[\A]  [ \bm{\nabla} \times \A^a(\x)]^2 {{ \Psi_0^2[\A] }\over {\langle 0|0\rangle}}
\nonumber \\
& = &  {\cal N} \int {{d\k}\over {(2\pi)^3}} { {\k^2} \over {2\Omega(|\k|)}},
\end{eqnarray}
where ${\cal N} = 2 \times (N^2_c - 1) \times {\cal V}$  counts the total (infinite) number of gluon 
degrees of freedom in volume ${\cal V}$ and $\Omega$ is the instantaneous gluon-gluon correlation function, 

\begin{equation} 
 \int D\A  {\cal J}[\A] \A^a(\p) \A^b(\q) 
{{ \Psi_0^2[\A]} \over {\langle 0|0\rangle}}  = {{\delta_{ab} } \over {2\Omega(|\p|)} } 
 (2\pi)^3 \delta(\p+\q). \label{Omega} 
\end{equation}
In the limit ${\cal J} \to 1$, $\Omega$ becomes equal to the gap function $\omega$~\cite{Reinhardt:2004mm,Szczepaniak:2003ve}. Evaluation of functional  integrals over non-gaussian distributions, like the one  in Eq.~(\ref{Omega}) for ${\cal J} \ne 1$ can be performed to  the leading order in $N_C$ by summing all planar diagrams. This produces a set of coupled integral (Dyson) equations for functions like $\Omega(p)$.  The Dyson equations contain, in general, UV divergencies.  To illustrate how renormalization takes place, let us consider expectation value of the inverse of the FP operator, 
\begin{equation} 
\delta_{ab} d(\x - \y) \equiv  \int D\A  {\cal J}[\A] g  (1-\lambda)^{-1}(\x,a;\y,b) 
 {{ \Psi_0^2[\A] } \over {\langle 0|0\rangle} }.  \label{d} 
\end{equation}
From translational invariance of the vacuum, it follows that the integral depends on $\x - \y$ and the Dyson equation for $d$ becomes simple  in momentum space. Defining, 
 $d(\x - \y) \to d(\p)  = \int d\x \exp(-i\k\cdot \x) d(\x)$, one obtains,  ($p = |\p|$, {\it etc.} ), 
 \begin{equation}
 {1\over {d(p)}} = {1\over {g(\Lambda)}}  -  {{N_C}\over 2} \int^\Lambda  {{d\q} \over {(2\pi)^3}}  {{(1 - \hat\q\cdot\hat\p)}  \over {\Omega(|\p-\q|) \q^2 } }  d(q). \label{eqd}
 \end{equation}
As expected from asymptotic freedom,  for large momenta, $\Omega(k)/k \to 1 
+ O(\log k)$;  the integral in Eq.~\ref{d}  becomes divergent as $q \to \infty$, and we need to introduce an UV cutoff $\Lambda$.  The cut-off dependence can, however, be removed by renormalizing the coupling constant $g \to g(\Lambda)$. The final equation for $d(p)$, renormalized at a  finite scale $\mu$, is obtained by subtracting from Eq.~(\ref{d}) the same equation evaluated at $p = \mu$.

One  also  finds that the expectation  value of  $(1-\lambda)^2$, which enters in  the Coulomb kernel, $K[\A]$, requires a multiplicative renormalization. We define the Coulomb potential as, 
\begin{equation}
 \int D\A  {\cal J}[\A]  K[\A](\x,a;\y,b) {{\Psi_0^2[\A] } \over {\langle 0|0\rangle}} 
  \equiv - \delta_{ab} V_C(\x - \y),  \label{V} 
\end{equation}
 and introduce a function  $f$ by,  
\begin{equation}
V_C(k) = \int d\x e^{ i\k\cdot\x} V_C(\x) \equiv - { {f(k) d^2(k) }  \over {k^2} },\label{Vk}
\end{equation}
This function then satisfies a renormalized Dyson equation,  
\begin{eqnarray}
f(k) & = &  f(\mu)  + \nonumber \\
& + &  \left[ {{N_C}\over 2}  \int {{d\q} \over {(2\pi)^3}}  {{(1 - \hat\q\cdot\hat\p)d^2(q)f(q)}  \over {\Omega(|\p-\q|) \q^2 } }   - (k \to \mu) \right].   \nonumber \\
\end{eqnarray}
Finally, the bare gap equation, $\delta [\langle 0|H_0|0\rangle/\langle 0|0\rangle]/\delta \omega(k) = 0$, contains a quadratic divergence proportional to  $\sim \Lambda^2$. This divergence is eliminated by  a single   relevant operator from the regularized Hamiltonian, the gluon mass term, which is  proportional to $\Lambda^2 \int d\x \A^a(\x)$.  The  renormalized gap equation determines the gap function $\omega(k)$, and it depends on a single dimensional subtraction constant, $\omega(\mu)$. 

The functions described above completely specify the variational ground state, and the  complete Fock space basis can be constructed by applying to this variational ground state  quasi-particle creation operators,  $\alpha^{a,\dag}(\k,\lambda)$,  defined by, 

\begin{eqnarray}
\A^a(\x) & =  & \int {{d\k} \over {(2\pi)^3}} {1\over {\sqrt{2\omega(k)}}} \left[ \alpha^a(\k,\lambda) \bm{\epsilon}(\k,\lambda) \right. \nonumber \\
 & &  \left. + \alpha^{a,\dag}(-\k,\lambda) \bm{\epsilon}(-\k,\lambda) \right ] e^{i\k\cdot\x}, \nonumber \\
 \bm{\Pi}^a(\x) & = & -i \int {{d\k} \over {(2\pi)^3}}  \sqrt{ {\omega(k)}  \over 2} 
\left[ \alpha^a(\k,\lambda) \bm{\epsilon}(\k,\lambda) \right. \nonumber \\
 & &  \left. - \alpha^{a,\dag}(-\k,\lambda) \bm{\epsilon}(-\k,\lambda) \right ] e^{i\k\cdot\x}. \nonumber \\
 \end{eqnarray}
 Here $\bm{\epsilon}$ represent  helicity vectors with $\lambda =\pm 1$.  
 This Fock space and  the corresponding  Hamiltonian matrix elements depend on four parameters (renromalization constants), $\omega(\mu)$, $d(\mu)$, $f(\mu)$ and one constant needed to regulate the FP determinant. The FP determinant enters into the Dyson equation for $\Omega(k)$.

 In principle, if the entire Fock space is used in building the Hamiltonian matrix and no approximations are made in diagonalization, the physical spectrum will depend on the single parameter of the theory {\it i.e} the renormalized coupling (or $d(\mu)$, {\it cf} Eq.~(\ref{eqd})). In practical calculations,  the Fock space  is truncated and this may introduce other renormalization constants.  Goodness of a particular basis,  for example the one built on the  state given in  Eq.~(\ref{psi0}), can be assessed by studying sensitivity of physical observables to  these residual parameters.

\begin{figure}
\includegraphics[width=3in]{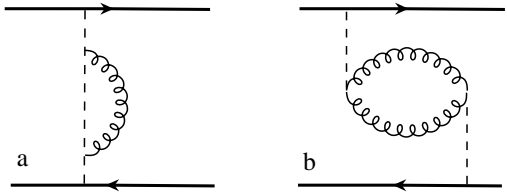}
\caption{\label{llogfig} The $O(g^4)$, one loop diagrams contributing to the leading log term in the expansion of the $\beta$-function in YM theory  with heavy sources. a) anti-screening dressing of the Coulomb potential by transverse gluons, b) Debye screening of the Coulomb potential by transverse gluons. The Coulomb potential is represented by the dashed line, and sources by thick lines. }
 \end{figure}

For example, if we define the running coupling as, $\alpha(k) \equiv f(k)d^2(k)$, so that   $V_C(k) = -{{4\pi \alpha(k)} \over {k^2}} $, we will find  that for large $k$, $\alpha(k) \propto (1/\log^{c}(k))[1 + O(1/\log(k))]$ where $c \sim 1.5$, \cite{Szczepaniak:2001rg}, while in full QCD the leading log has power  $c=1$. The discrepancy arises  because we used the single Fock state, $|0\rangle$ in definition of $V_C$ (and $\alpha$). This omits, for example, the contribution from the two-gluon Fock state, as shown in 
 Fig.~\ref{llogfig}. This  two gluon intermediate state clearly impacts the short range behavior of  the Coulomb interaction, but, as discussed in ~\cite{Szczepaniak:2001rg}, it  is not expected to affect the long range part (partially because the low momentum gluons develop a large constituent mass). Similarly, in ~\cite{Szczepaniak:2003ve}, the role of the FP determinant has been analyzed, and it was shown that it does not make a quantitative difference leading to $\Omega(p) \sim \omega(p)$.  

   This is in contrast, however,  to the results of ~\cite{Feuchter:2004mk}. We think this discrepancy originates from the difference in the boundary conditions which in~\cite{Feuchter:2004mk} 
lead to  $f(k)=1$. This makes possible for the gap equation to have a solution   for  $\omega(k)$  which rises at low momenta. If $f(k) \ne 1$ and, in particular, if $f(k)$ grows as $k\to 0$,    which is necessary if  $V_C(R)$ is to grow linearly for large $R$, we find that $\omega(k)$ has to be finite as $k\to 0 $. A more quantitative  comparison is currently being pursued. We also note that lattice simulations~\cite{Langfeld:2004qs} are consistent with the results of ~\cite{Szczepaniak:2003ve,Szczepaniak:2001rg}.   

 In the following, we will thus set ${\cal J} =1 $, which makes $\Omega=\omega$,  and use the solutions   for $f(k)$, $d(k)$ and $\omega(k)$ found in  Ref.~\cite{Szczepaniak:2001rg}.  

Finally, we want to stress that the Coulomb potential, defined in Eqs.~(\ref{V}),~(\ref{Vk}),  gives   the  energy  expectation value in the state obtained by adding the  $\qq$  pair to the vacuum of Eqs.~(\ref{0}),~(\ref{psi0}), {\it i.e},   
  \begin{equation} 
  \langle \qq |H|\qq\rangle =  C_F V_C(R) - C_F V_C(0),
  \end{equation}
 with $C_F V_C(0)$ originating from self-energies,    
 and 
 \begin{eqnarray}
  |\qq \rangle &= & |R,N=0,\Sigma^+_g\rangle  \nonumber \\ 
  & = &  {1\over {\sqrt{N_C}}} h^{\dag}\left({R\over 2} \hat\z\right)\eta^{\dag}\left(-{R\over 2}\hat\z\right)  {{| 0\rangle } \over {\langle 0|0\rangle}}. 
 \label{qq0} 
  \end{eqnarray} 
The state $|R,N=0,\Sigma^+_g\rangle$ refers the the ground state $(N=0)$ with spin-partiy quantum numbers $\Lambda^Y_{PC} = 0(\Sigma)^+_g$.  The energy $C_F V_C(R)$ should be distinguished from $E_0(R)$ in Eq.~(\ref{qq}). The latter is evaluated using the {\it true} ground state of the $\qq$ system while the former is evaluated in a  state obtained by simply adding a $\qq$ pair to the vacuum. Since a $\qq$ pair is expected to polarize the gluon distribution ,these two states are different.  Furthermore, in this work, the $|\qq\rangle$ state in Eq.~(\ref{qq0}) is obtained by adding the $\qq$ pair to the {\it variational} state  of the vacuum and not to the {\it true} vacuum state in the absence of sources.

  \subsection{Fitting the Coulomb Potential} 
As discussed above,  the Coulomb energy, $C_F V_C(R)$,  represents the expectation value of the Hamiltonian in a particular $\qq$ state (given in Eq.~(\ref{qq0})), which is not the same as the true eigenstate of the Hamiltonian for the $\qq$ system  as defined in Eq.~(\ref{qq}). The latter has energy $E_0(R)$. 
 
According to ~\cite{Zwanziger:2002sh}, $C_F V_C(R) > E_0(R)$ and numerical results in ~\cite{Greensite:2004ke} further indicate  that  for large $R$, $C_F V_C(R) \sim \sigma_C R $ and $E_0(R) \sim \sigma R$ with the Coulomb string tension, $\sigma_C$, being approximately three times larger then $\sigma$.  In ~\cite{Szczepaniak:2001rg} we, however,  fitted $d(\mu)$, $f(\mu)$ and $\omega(\mu)$ so that  $C_F V_C(R) \to E_0(R)$, and  a number of phenomenological studies have been successful with those parameters ~\cite{Adler:1984ri,Szczepaniak:1995cw,Ligterink:2003hd,Szczepaniak:2003mr}.  It should be noted, however, that the results from ~\cite{Greensite:2004ke} for $C_F V_C(R)$ may not directly apply to our analysis since  the  $\qq$ state used here  to define $V_C(R)$  may be different from the one used in lattice computations of $V_C(R)$. 
      Guided by the successes of the phenomenological applications of our approach we 
  proceed with fitting  $C_F V_C(R)$ to $E_0(R)$. It is clear, however, 
 that since the $\qq$ state of Eq.~(\ref{qq0}) is a variational state, $C_F V_(R)$ should be greater than  $E_0(R)$~\cite{Zwanziger:2002sh}. We will nevertheless proceed with the approximation $C_F V_C(R) = E_0(R)$   and examine the consequences afterwards.  
   
In ~\cite{Szczepaniak:2001rg}, we have found that  the numerical solutions to the set of coupled Dyson equations for $d(k)$, $f(k)$ and $\omega(k)$ can be well represented by,   
  
\begin{equation}
d(k) =    \left\{ \begin{array}{cc} 3.5 \left( {m_g \over k} \right)^{0.48} &  \mbox{ for } k  <  m_g \\
 3.5 \left( {{ \log(2.41) } \over {\log(k^2/m_g^2 + 1.41)} } \right)^{0.4}  & \mbox{ for } k > m_g ,
 \end{array}\right.
 \end{equation}
 \begin{equation}
 f(k) = \left\{ \begin{array}{cc} 1.41 \left( {m_g \over k} \right)^{0.97}  & \mbox{ for } k < m_g \\
  1.41 \left( {{ \log(1.82) } \over {\log(k^2/m_g^2 + 0.82) } } \right)^{0.62} & \mbox{ for } k > m_g ,
  \end{array} \right.
  \end{equation} 
  \begin{equation}
  \omega(k) = \left\{ \begin{array}{cc} m_g  & \mbox{ for } k < m_g \\
   k  & \mbox{ for } k > m_g . \end{array} \right.
   \end{equation} 
The parameter $m_g = 650 \mbox{ MeV}$  effectively represents the constituent gluon mass. It should be noticed, however, that $\omega(k)$ is the gap function and not the single quasi-particle energy. This energy, denoted by  $E_g(k)$  is given by,  

   \begin{eqnarray}
 &&   E_g(k)  =   \omega(k)\left[ 1 
 - {N_C \over 2} \int {{d\q} \over {(2\pi)^3}} V_C(\k- \q) {{1 + \hat\k\cdot\hat\q} \over {2\omega(q)}}
     \right].   \nonumber \\
     & & \label{eg}
   \end{eqnarray}
Since $V_C(k) = -f(k)d^2(k)/k^2$, which for small $k$ grows faster then $k^3$, the integral in 
<     Eq.~(\ref{eg}) is divergent. This IR divergence is a manifestation of the long range nature  of the confining Coulomb  potential which removes single, colored excitations from the spectrum. As will be explicit in the examples studied later, residual interactions between colored constituents in color neutral states cancel such divergencies and result in a finite spectrum for color neutral states.  In the following analysis, we will also need the Coulomb potential in coordinate space. We find it practical to approximate the numerical FT of $V_C(\k-\q)$ by,

 \begin{figure}
\includegraphics[width=2.6in,angle=270]{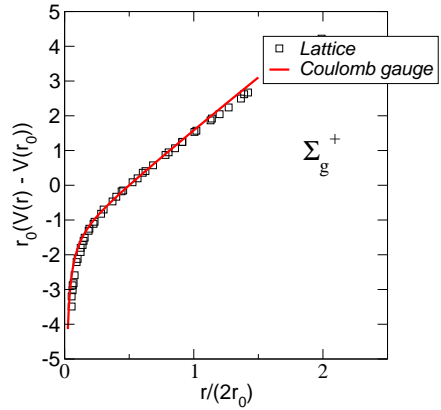}
\caption{\label{coul} Comparison between $V(R) = C_F V_C(r)$ from Eq.~(\ref{vcr}) (solid line)  and  
 $V(R) = E_0(R)$ lattice data from ~\cite{Juge:1997nc} ($r_0 = 1/450 \mbox{ MeV}^{-1}$).}
 \end{figure}

\begin{equation}
V_C(r) =  b r - {\alpha \over {r \log^c[ (r \Lambda)^{-1} + a]} }, \label{vcr}
\end{equation}
with  $b= 0.20 \mbox{ GeV}^2$, $\alpha=0.83$, $\Lambda = 0.63\mbox{ GeV}$, $a=1.24$ and $c=1.51$.
Comparison between $C_F V_C(R)$ and $E_0(R)$ obtained from lattice computations is shown in Fig.~\ref{coul}. 

We now proceed to the main subject of this paper, namely to  investigate the difference between $E_0(R)$ computed using the single Fock space approximation to the $\qq$ state ({\it i.e} without modification of the gluon distribution) and the solution of Eq.~(\ref{qq}) which accounts for modifications in the gluon distribution in the vacuum in presence of $\qq$ sources. We will also compute the first excited potential with the $\Sigma^+_g$ symmetry.

\section{Adiabatic potentials} 

To diagonalize the full Hamiltonian in the Fock space described above, in principle,  requires an infinite number of states. In the zeroth-order approximation, $E_0(R) = C_FV_C(R)$, a single state with no 
<  quasi-gluons was used. At vanishing $\qq$ separation, we expect the wave function of the system to be identical to that of the vacuum, and the approximation becomes exact.  One also expects that the average number of quasi-gluon excitations in the full wave functional of Eq.~(\ref{qqwf}) increases with the $\qq$ separation.  We thus start by examining the approximation based on adding a single quasi-gluon and truncate the Hamiltonian matrix to a space  containing 
 $|\qq\rangle $ and $|\qq g\rangle$  states,

\begin{eqnarray}
& & \left[\begin{array}{cc}  \langle \qq | H \qq  \rangle & \langle \qq|H| \qq g\rangle \\
 \langle \qq g|H |\qq \rangle & \langle \qq g|H | \qq g \rangle \end{array} \right] 
 \left[ \begin{array}{c} |\qq \rangle \\ |\qq g \rangle \end{array} \right]  \nonumber \\
  & &    = E_N(R) \left[\begin{array}{c} |\qq \rangle \\ |\qq g \rangle \end{array} \right] .  \label{mix} 
 \end{eqnarray}
The $|\qq\rangle $ state is given in Eq.~(\ref{qq0}).  In the quasi-particle representation the state with a single gluon and $\Lambda^Y_{PC}$ quantum numbers, $|\qq g\rangle = |R,n,\Lambda^Y_{PC}\rangle$  is given by,  
 \begin{eqnarray}
& &  |R,N,\Lambda^Y_{PC} \rangle  =    \sum_{j_g,\xi,\mu,\lambda} \sqrt{ {2j+1}\over {8\pi C_F N_C }} 
   \int {{d\k} \over {(2\pi)^3}}  \nonumber \\
   & &  \left[ D^{j_g*}_{\Lambda\mu}(\hat\k) 
 + \eta_Y  D^{j_g*}_{-\Lambda\mu}(\hat\k)  \right]
 \psi^{j_g}_{N}(k) \chi^{\xi}_{\mu\lambda} |R,\k,\lambda\rangle, \nonumber \\
  \label{wfg} 
    \end{eqnarray}
  for $\Lambda \ne 0$ and,
   \begin{eqnarray}
& &  |R,N,0^{PC}_Y \rangle =    \sum_{j_g,\xi,\mu,\lambda} \sqrt{ {2j+1}\over {4\pi C_F N_C }} 
\nonumber \\ 
& \times &   \int {{d\k} \over {(2\pi)^3}}  D^{j_g*}_{0\mu}(\hat\k) 
 \psi^{j_g}_{N}(k) \chi^{\xi}_{\mu\lambda} |R,\k,\lambda\rangle, \nonumber \\
  \label{wfg1} 
    \end{eqnarray}
for $\Lambda=0$ ($\Sigma$ potentials) 
      where 
  \begin{equation}
  |R,\k,\lambda\rangle = h^{\dag}\left({R\over 2}\hat\z\right) \alpha^{\dag}(\k,\lambda) \eta^{\dag}\left(-{R\over 2} \hat\z\right) {{|0 \rangle} \over {\langle 0|0\rangle} }, 
  \end{equation}
and $\alpha^{\dag} = \alpha^{a,\dag} T^a $. In Eqs.~(\ref{wfg}),(\ref{wfg1}), $j_g$ is the total angular momentum of the quasi-gluon. For vanishing separation between the quarks, the system has full rotational symmetry, and $j_g$ becomes a good quantum number. In general, the system is invariant only under rotations around the $\qq$ axis. It is only the projection of the total angular momentum, $\Lambda$, that is conserved and states with different $j_g$ become mixed.  The wave function $\chi^\xi_{\mu\lambda}$ represents the two possibilities for the spin-oribt coupling of given parity, ($j_g = L_g$ or $j_g = L_g \pm 1$). It is given by
$\delta_{\mu\lambda}/\sqrt{2}$ for $\xi= 1$ and  $\lambda\delta_{\mu\lambda}/\sqrt{2}$ for $\xi=-1$, corresponding to TM (natural parity) and TE (unnatural parity) gluons, respectively. 
 Finally $\eta_Y$ determines the behavior under reflections in the plane containing the $\qq$ axis, {\it i.e.}, the $Y$-parity. 
 
 The radial wave functions, $\psi^{j_g}_N(k)$, labeled by the radial quantum number $N$ and $j_g$, are obtained by  diagonalizing  the  full Hamiltonian in the Fock space spanned by the $\qq g$ states alone, {\it i.e} by solving the equation, 
\begin{equation}
P H  P |R,N, \Lambda^Y_{PC} \rangle = V^{qqg}_{C,N}(R) |R,N,\Lambda^Y_{PC} \rangle. \label{hgbare}
\end{equation}
 Here $P$ projects on the $|\qq g\rangle$ state and $V^{qqg}_{C,N}(R)$ are the {\it bare} 
 energies of the excited adiabatic potentials, {\it i.e.}, without mixing between  states with a different number   of quasi-gluons. Analogously, $C_F V_C(R)$ is the {\it bare}  
 ground state energy $E_0(R)$.   The matrix elements of $PHP$ are shown in Fig.~\ref{vqqg} and given explicitly in the Appendix.

 \begin{figure}
\includegraphics[width=3in]{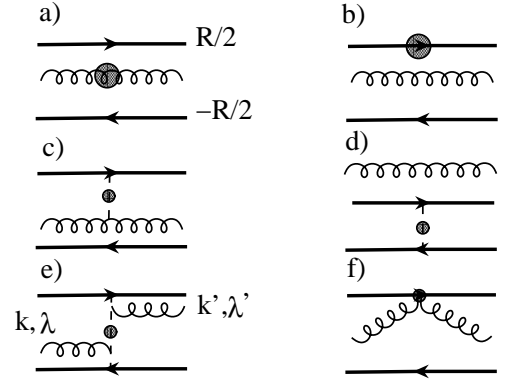}
\caption{\label{vqqg} Matrix elements, $\langle R, \k',\lambda' | H| R,\k, \lambda \rangle$. Diagrams a) and b) represent gluon and quark self energies, respectively. Diagrams c) and d) represent the Coulomb interaction, $V_C$ between the gluon and one of the quarks and between the two quarks, respectively.  In the bottom row, diagrams e) and f)  describe matrix elements of the interaction term resulting from expansion of the Coulomb kernel $K[A]$ in up to one power in  gluon field. }
 \end{figure}

 The mixing matrix element,
\begin{equation}
\langle \qq| H |\qq g \rangle = V^{qq,qqg}_{C,N}(R),
\end{equation}
depends on the number of  bare, $\qq g$ states  from Eq.~(\ref{hgbare}) kept,  $N = 1,\cdots N_{max}$ and the separation between the sources, $R$.  It is shown in Fig.~\ref{vmix} and given in the  Appendix.

 \begin{figure}
\includegraphics[width=2.5in]{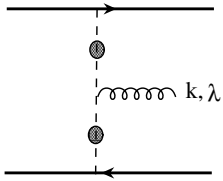}
\caption{\label{vmix} The matrix element $\langle R |H|R,\k,\lambda\rangle$. It originates from  expansion of the Coulomb kernel $K[A]$ to first order in $A$} 
 \end{figure}

The $(N_{max} +1) \times (N_{max} + 1)$  Hamiltonian matrix shown in  Eq.~(\ref{mix}) 
  is explicitly given by, 
 
\begin{equation} 
H_{NM} = \left\{ \begin{array}{ll} C_F\left[ V_C(R) - V_C(0) \right] &  N=M=0 \\
         V^{qq,qqg}_{C,M}(R) &   N=0, M=1-N_{max} \\
          {V^{qq,qqg}}^*_{C,N}(R)  &   N=1-N_{max}, M= 0 \\
          V^{qqg}_{C,N}(R)\delta_{NM}  & N,M = 1-N_{max} 
          \end{array} \right. . \label{mixmat} 
          \end{equation}

\section{ Numerical  Results} 
In terms of $\xi$ and $\eta$, the $PC$ and $Y$ quantum numbers  of the gluonic field are given by, 
 \begin{equation}
 PC = \xi (-1)^{j_g + 1}, \; Y= \left\{ \begin{array}{c} \xi \eta_Y (-1)^\Lambda  \mbox{ for } \Lambda \ne 0 
 \\ \xi \mbox{ for } \Lambda = 0 \end{array} \right.  .  \label{PC} 
  \end{equation} 
 In the following, we will concentrate on the  states with $\Lambda=0$, $PC=g(+)$ and $Y=+$,  {\it i.e.}, of $\Sigma^+_g$ symmetry, since it is only these states that mix the bare $|\qq\rangle$  state with the states with non-vanishing number of gluons.   

\begin{figure}
\includegraphics[width=2.6in,angle=270]{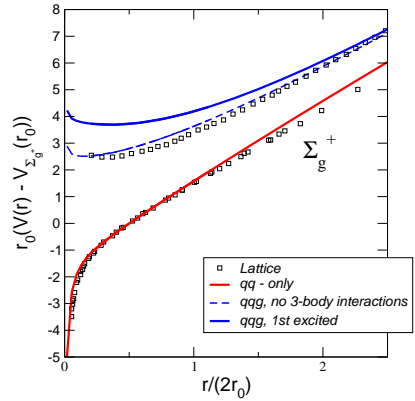}
\caption{\label{sigma_no_mix_new} Comparison between $V(R) = C_F V_C(r)$ from Eq.~(\ref{vcr}) (solid line)  and the 
 $V(R) = E_0(R)$ lattice data from ~\cite{Juge:1997nc} ($r_0 = 1/450 \mbox{ MeV}^{-1}$).}
 \end{figure}

For the $\Sigma^+_g$ potentials, the wave function contains TM gluons, $\xi = 1$ of natural parity and $PC=+1$ which implies $j_g = 1,3,\cdots$.  As discussed above, for $R\to0$, $j_g$ becomes a good quantum number, and we have verified numerically that for $R$ in the range considered here the contributions from $j_g = 3$ and higher are at  a level of a few percent. Diagonalization  of the Hamiltonian in the $\qq g$ subspace alone, leads to the $V^{\qq g}_{C,N}(R)$ potential which is shown in Fig.~\ref{sigma_no_mix_new} (upper solid line) for the lowest excitation with  $N=1$. The dashed line is the result of using the one- and two-body interactions depicted in Figs.~\ref{vqqg}a-d.  ($H_{\ref{vqqg}a}-H_{\ref{vqqg}d}$ in Eq.~(\ref{htot})). These are also the interactions that were used in ~\cite{Swanson:1998kx}.  When the three-body interactions shown in Fig.~\ref{vqqg}e,f are added, the energy moves up.  This discrepancy is then also a measure of how far our variational, truncated Fock space expansion is from the true excited state.  The three-body potential is expected to be responsible   for reversing the ordering between  the $\Pi_u$ and $\Pi_g$ surfaces; with only one- and two-body interactions, the $\Pi_g$ potential has lower energy than $\Pi_u$, which is inconsistent with the lattice data~\cite{Swanson:1998kx}. In the Appendix, we also show that the three-body term is suppressed at large separations, and thus the net  potential approaches the Casimir scaling $C_F b R$ limit as $R \to \infty$.     Finally, we note that when the Fock space is restricted to single quasi-gluon excitations, the  diagrams   in Fig.~\ref{vqqg} and Fig.~\ref{mix} represent the complete set of Hamiltonian matrix elements .

 The general features of higher excitations, $V^{qqg}_{C,N}(R)$ for $N>1$,  follow from the structure of the Hamiltonian, which  represents  a one-body Schr{\" o}dinger equation for the single quasi-gluon wave function in momentum space. The kinetic energy corresponds to the one-body diagram in  Fig.~\ref{vqqg}a and the potential to the diagrams in Fig.~\ref{vqqg}c,e,f. 
The diagrams in  Figs.~\ref{vqqg}b,d give an  $R$-dependent shift describing the $\qq$ self-interactions and $\qq$ octet potential. The IR singularity  in the gluon kinetic energy, $E_g$, is canceled by the collinear singularity of the two-body potential, the $\qq$ self energy and $\qq$  octet potential.
 On average, gluon  kinetic energy  contributes an effective quasi-gluon mass of the order of  $m_g$. Quasi-gluon are thus heavy, and adding Fock space components, with more gluons, $|\qq, 2g\rangle,  \cdots |\qq n_g g \rangle$, for small-$R$ will result in  higher adiabatic potentials with $(N=2,3\cdots$)  that are split from the first excited state by $\sim n_g m_g$. At large $R$, the  two-body  Coulomb potential dominates and together with Coulomb energies of the pair-wise gluon interactions,  results in the Casimir scaling (we will discuss this in more detail in the following section).   In the absence of mixing between Fock space components the number of quasi-particle gluons in the $|\qq, n_g g\rangle$ state  is conserved, and they directly map in to the tower of excited adiabatic potentials.

 \begin{figure}
\includegraphics[width=2.6in,angle=270]{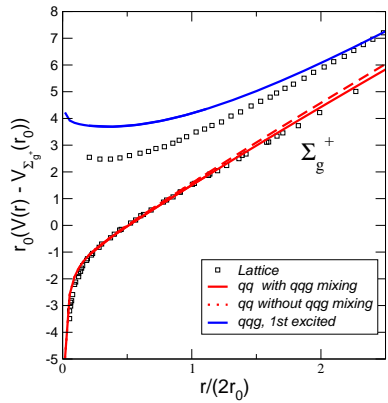}
\caption{\label{sigma_with_mix_new-2} Comparison between $V(R) = C_F V_C(r)$ from Eq.~(\ref{vcr}) (solid line)  and the 
 $V(R) = E_0(R)$ lattice data from ~\cite{Juge:1997nc} ($r_0 = 1/450 \mbox{ MeV}^{-1}$).}
 \end{figure}
 
\begin{figure}
\includegraphics[width=2.6in,angle=270]{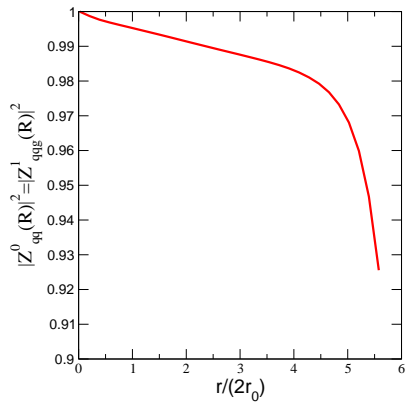}
\caption{\label{prob} Normalized probability of finding the bare $|\qq\rangle$ state in the full ground state of the $|qq,N=0\rangle$ (which is also equal to the probability of finding the $|\qq g\rangle$ state in the first excited $|qq, N=1\rangle$ state).}
 \end{figure}

 We will now address the effects of mixing between $|\qq\rangle$ and $|\qq g\rangle$ states.  The only non-vanishing diagram is shown in Fig.~\ref{mix}. Since, as discussed above,  the $V^{qqg}_{C,N}(R)$ potentials are split from the first excited state, $N=1$, by at least $m_g$, the mixing matrix   in Eq.~(\ref{mixmat}) saturates quickly, and in practice, only the $N=1$ state is relevant. However, even this single state mixing leads to a very small energy shift. In Fig.~\ref{sigma_with_mix_new-2} the dashed  line corresponds to the energy of the ground state without mixing, (the same as the solid line in Fig.~\ref{sigma_no_mix_new}), and the solid line shows the effect of mixing. The effect of the mixing is small. Numerically, we find that the full ground state, 
 \begin{equation}
 |\qq, N=0\rangle = Z^0_{qq}(R) |\qq\rangle  +  Z^0_{qqg} |\qq g\rangle, 
 \end{equation}
is still dominated by the $|\qq \rangle$ component and the first excited 
state, 
\begin{equation}
 |\qq, N=1\rangle = Z^1_{qq}(R) |\qq\rangle  +  Z^1_{qqg} |\qq g\rangle, 
 \end{equation}
 by the $|\qq g \rangle$ component. The probabilities of each are shown in Fig.~\ref{prob}. 
 We see that, for distances between sources as large as $5\mbox{ fm}$, the
  admixture of the gluon  component  is only of the order of $10\%$.

 This small admixture of the $|\qq,g\rangle$ in the full ground state is  correlated with the small shift in the $\Sigma^+_g$ surface  shown in 
 Fig.~\ref{sigma_with_mix_new-2} and would justify  using the ground state, exact $\Sigma^+_g$ energy to constrain the Coulomb potential $V_C$. This is, however,  contradicting the results of Ref.~\cite{Greensite:2003xf} where the effect of mixing must be large since it results in a factor of three in the ratio of the unmixed to mixed string tensions. One possible explanation is that there is an accidental suppression of the mixing interaction matrix element for the two states considered here, $|\qq\rangle$ and $|\qq g\rangle$. Inspecting Eq.~(\ref{hvmix}), we note that due to the gradient coupling of the transverse gluon to the Coulomb line, the coupling vanishes both for small and large-R.  In contrast, a two gluon state can be coupled to $|\qq\rangle$ with either the Coulomb line mediated interaction as shown in Fig.~\ref{higher}a or the quark density- gluon density interaction shown in Fig.~\ref{higher}b. As discussed in the Appendix, at large distances the former is suppressed and it is easy to show that the latter is proportional to $C_F V_C(\x - \R/2) +  C_F V_C(\x + \R/2)$ (once the gluon spin is neglected) and persists at large distances. In the large-$N_C$ limit $C_F  = N_C/2 (1 + O(1/N_C))$.  It is therefore possible that the $|\qq, 2g\rangle$ component of the full $|\qq, N=0\rangle$ state is actually   more important then the $|\qq g\rangle$ one. We will investigate this further in section~\ref{chainsec}.

  \begin{figure}
\includegraphics[width=2.5in]{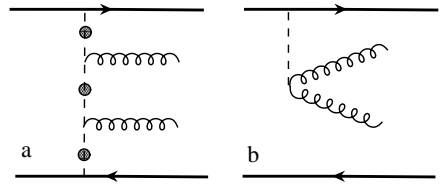}
\caption{\label{higher}  Matrix elements $\langle qq|H|\qq,2g\rangle$ leading to the $|\qq, 2g\rangle$ component in the ground state $\Sigma^+_g$ potential, a) interaction mediated via the Coulomb line coupled to quark sources, b) interaction between a single quark and the gluon charge density.  }
 \end{figure}

\subsection{\label{chainsec} Multi-gluons states and the chain model} 

As shown above, the quasi-gluon degrees of freedom defined in terms of a variational quasi-particlue vacuum provide an attractive basis for describing gluon excitations. This is in the sense that for source separations relevant for phenomenology the color singlet states can be effectively classified in terms of the number of quasi-gluons. This basis, however, does overestimate the energies (as expected in a variational approach), and this fact together with lessons from other models can give us guidance for how to improve on the variational state of the $\qq$ system.  As the separation between quarks increases one expects the average number of gluons in the energy eigenstate to increase. This is because it becomes energetically favorable to add a constituent gluon which effectively screens the $\qq$ charge.  Furthermore, the spacial distribution of these gluons is expected to be concentrated near the $\qq$ axis  in order  for the energy distribution to be that of a flux tube, as measured by the lattice. An improvement in the ansatz wave functional will therefore result in a more complicated Fock space decomposition with a large number of quasi-gluons present, even at relatively small separations between the sources. In this section we will first discuss how multi-gluon states indeed become important, even in the case of the quasi-gluon basis used here. We then compare with expectations from other models and discuss the possible directions for improving the quasi-gluon basis.

 \begin{figure}
\includegraphics[width=2.5in]{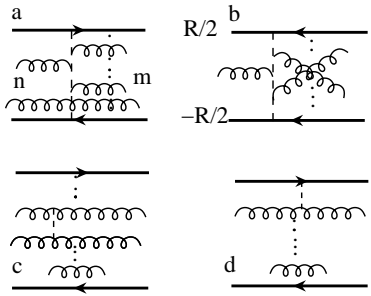}
\caption{\label{multi}  Typical diagrams contribution to mixing between $n$ and $m$ gluon states. 
Vertical dots represent any number of gluons not affected by the interaction. a) mixing mediated by the Coulomb potential, b) same as in a) with rearrangement of gluons, c) long-range Coulomb interaction between gluon charge densities, d) same as in c) but  with the charge density of the quark sources. } 
 \end{figure}

As discussed in the Appendix, at large separations the interactions between multi-gluon Fock states mediated by the Coulomb potential, shown in Fig.~\ref{multi}a,b, require all but two gluons to be at relative separations smaller than $R$. Furthermore, rearrangement of gluons leads to $1/N_C$ suppression.  For large $R$, the largest diagonal matrix elements of $H$ are the ones corresponding to   the long-range Coulomb interaction between charge densities as shown in Fig.~\ref{multi}c,d. To leading order in $N_C$, the gluons should be path-order along the $\qq$ axis. For simplicity, we will neglect the gluon spin and use a single wave function to represent a state with an arbitrary number of gluons. We write

\begin{figure}
\includegraphics[width=2.6in,angle=270]{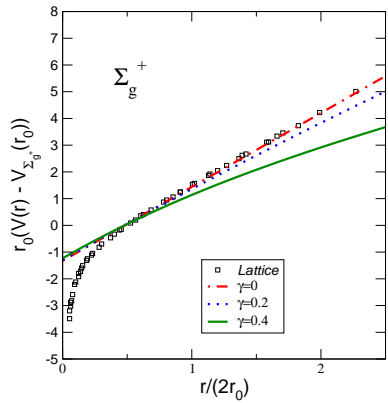}
\caption{\label{model}  Shift in the ground state $\Sigma^+_g$ energy due to coupling with muti-gluon states of the model Hamiltonian of Eqs.~(\ref{mod-1}),~(\ref{mod-2}). The maximum number of states was taken to be $n_{g, max} = 40$. The other parameters are $b=0.21\mbox{ GeV}^{-2}$ and $m_g = 0.65 \mbox{ GeV}$. } 
 \end{figure}

\begin{figure}
\includegraphics[width=2.6in,angle=270]{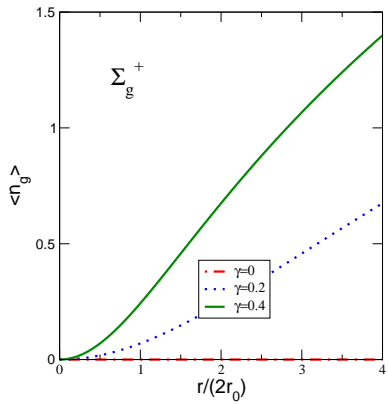}
\caption{\label{ng}  Average number of quasi-gluons in the full eigenstate of the model Hamiltonian of     Eqs.~(\ref{mod-1}),~(\ref{mod-2}). } 
 \end{figure}

 \begin{widetext} 
 \begin{equation}
 |\qq, n_g g\rangle = N_{n_g}  \int_{-R/2}^{R/2} dx_{n_g} \alpha^{\dag}(x_{n_g}) \int_{-R/2}^{x_{n_g}} dx_{n_g-1} \alpha^{\dag}(x_{n_g-1}) \cdots  \int_{-R/2}^{x_3} dx_2 \alpha^{\dag}(x_2)    \int_{-R/2}^{x_2} dx_1 \alpha^{\dag}(x_1) |0\rangle  \label{chain}
\end{equation}
\end{widetext} 
 where we have also forced all gluons to be on the $\qq$ axis. The  factor 
$N_{n_g} = (n_g !/C_F N_C R)^{1/2}$ is, to leading order in $N_C$ fixed by the normalization condition,  $\langle \qq, n_g g| \qq, n'_g g\rangle = \delta_{n_g,n'_g}$, where we used $[\alpha(x_i),\alpha^{\dag}(x_j)] = \delta_{ij}$.  In this basis, the diagonal matrix elements of the Hamiltonian ({\it cf.} Fig.~\ref{multi}c,d) add  up to 
  \begin{equation}
  H_{n_g n'_g } = \langle \qq, n_g g| H |\qq, n'_g g \rangle =  C_F V_C(R) \to C_F b R  \delta_{n_g,n'_g} .\label{mod-1}
   \end{equation}
The off-diagonal matrix elements are dominated by interactions between color charges, {\it e.g.}, similar to the ones in Fig.~\ref{higher}b, but with the upper vertex attached to a gluon line.  With the approximations leading to Eq.~(\ref{chain})  a vertex which either annihilates or creates two gluons results in a vanishing matrix element since in our basis no two gluons are at the same point. Smearing each gluon in the coordinate space by a distance of the order of $1/m_g$ will give a finite matrix element , which just like the diagonal matrix elements grows linearly with $R$, 

    \begin{eqnarray}
  H_{n_g n'_g }  & =  & \langle \qq, n_g g| H |\qq, n'_g g \rangle \nonumber \\
   & \to &   \gamma C_F b R \left[ \delta_{n_g,n'_g+2} +  \delta_{n_g,n'_g+2} \right] , \label{mod-2}
   \end{eqnarray}
where $\gamma$ is a parameter representing the effect of a smearing, 
 and we expect  $|\gamma| < O(1)$.  In addition, each gluon has a kinetic energy of the order of $m_g$, so $H_{nn} \to H_{nn} + n m_g$.  The model Hamiltonian can be easily  diagonalized  numerically, and  in Fig.~\ref{model}, we  plot the energy of the ground state and of the first excited state as a function of $R$.  It is clear that in the absence of accidental spin suppression, which, as discussed earlier, takes place for the $\langle \qq | H |\qq g\rangle$ mixing matrix, the effect of the mixing with two and more gluons can produce shifts in the lowest adiabatic potential and decrease the Coulomb string tension by as much as a factor of $\sim 2 $ at     $R\sim 3r_0 = 2.6 \mbox{ fm}$.  Finally, in Fig.~\ref{ng}  we plot the average number of gluons in the ground state of the model Hamiltonian. As expected, the number of gluons grows with $R$; however, still a small number of quasi-gluons contributes to the ground  state at these separations, which again provides justification for the quasi-gluon description.

\section{Summary and Outlook} 
We computed the ground state energy and the energy of the first excited $\qq$ potential with the $\Lambda^Y_{PC} = \Sigma^+_g$ symmetry.  We used the quasi-particle basis of constituent gluons based on a variational ground state to build the Fock space representaion. We found that the  $\qq$ state can be  well approximated by a superposition of the bare $\qq$ state and  a few quasi-gluons. The exact computation in which the bare $\qq$ sate mixes with a state containing a single quasi-gluon  leads to negligible change in the energy of the bare (Coulomb) $\qq$ system. We found that this is due to an accidental small mixing matrix element of the Coulomb gauge Hamiltonian. We have discussed the general properties of the mixing matrix between states with an arbitrary number of gluons, and  using a simple approximation, we have found a good agreement with the lattice data.  The lattice data indicates that there is a change in slope between the Coulomb and the true, Wilson potential~\cite{Greensite:2003xf}.   
  Based on the representation used here, we interpret this in terms of   quasi-gluon  excitations  rather than in terms of a flux-tube-like degrees of freedom.  We also note that lattice data on splitting between several excited  $\qq$ states does not unambiguously show a string-like behavior for separation as large as   $2-3\mbox{ fm}$~\cite{Juge:2002br}. In fact, the splittings are almost constant, 
   although why lattice data has such a behavior is not completely understood  (including a possible systematic error)~\cite{colin}.  In fact this data is consistent with the quasi-gluon picture where each quasi-particle adds kinetic energy of the order of the effective gluon mass. 
The full excitation spectrum  as well as distribution of energy density is currently being investigated.

\section{Acknowledgment}
I would like to thank J.~Greensite, C.~Morningstar and E.~Swanson, 
 for several discussions and C.~Halkyard for reading the manuscript.
  This work was supported in part by the US
Department of Energy grant under contract 
 DE-FG0287ER40365. 

\section{Appendix} 
Here we list matrix elements of the Hamiltonian in the basis spanned by $|\qq\rangle 
= |R,N=0,\Lambda^Y_{PC}\rangle$ and $|\qq g\rangle = |R,N\ne0, \Lambda^Y_{PC} \rangle$. 
 The $|\qq \rangle$ state  
exists only in the $\Lambda^Y_{PC}= \Sigma^+_g$ configuration. Thus mixing matrix elements are non-vanishing  for $|\qq g\rangle$ with  $\Sigma^+_g$ spin-parity quantum numbers only. 

For each $j_g$, the wave functions $\psi_{N,j_g}(k)$ are expanded in a complete  
  orthonrmal basis of functions $\phi_{m,j_g}(k)$ 
  \begin{equation}
  \psi_{N,j_g}(k) = \sum_{m=1}^{m_{max}}  a^{m}_{N,j_g} \phi_{m,j_g}(k)
  \end{equation} 
with normalization,  $\int {{dk k^2} \over {(2\pi)^3}}   \phi^*_{m',j'_g}(k) \phi_{m,j_g}(k) = \delta_{m',m}\delta_{j'_g,j_g}$.    The expansion coefficients are computed by diagonalizing the  $(m_{max}j_{g,max})  \times (m_{max}j_{g,max})$  matrix, $\tilde{H}_{m'j_g';m,j_g}$,   obtained by evaluating the diagrams in Fig.~\ref{vqqg},

\begin{equation}
\tilde{H_3}  = H_{3a} + H_{3b} + \cdots + H_{3f}, \label{htot}
\end{equation} 
evaluated in the basis of functions $\phi_{m,j_g}$. In numerical computations  for each $j_g$, we used a momentum grid as the basis functions. 
    The numerical results presented were for a single $j_g$  determined from 
    Eq.~(\ref{PC}) after verifying that increasing $j_g$  changes the computed spectrum by at most a few percent. 
 For arbitrary $\Lambda^Y_{PC}$ the Hamiltonian matrix elements are given by, 

\begin{equation} 
H_{3a} = {{\delta_{j'_g,j_g}}\over 2}  \int {{dk k^2} \over {(2\pi)^3}} \phi^*_{m',j_g}(k) E_g(k) \phi_{m,j_g}(k), 
\end{equation}

\begin{eqnarray}
H_{3b} & = &  -  C_F V_C(0)\delta_{m',m}\delta_{j'_g,j_g}  \nonumber \\
 &  = &  - 4\pi C_F \int {{dk k^2} \over {(2\pi)^3}} V_C(k)  \delta_{m',m}\delta_{j'_g,j_g}  ,
\end{eqnarray}
with 
\begin{equation}
V_C(k) = - {{d^2(k)f(k)} \over {k^2}},
\end{equation}
\begin{widetext}
\begin{eqnarray} 
H_{3c} & =  & 
{N_C \over 2} \sum_{\lambda,\lambda',\sigma,\sigma',\mu} \int {{d\q} \over {(2\pi)^3}} 
 \int {{d\k} \over {(2\pi)^3}} 
 \phi^*_{m',j'_g}(q) \phi_{m,j_g}(k)\int d\x  \left[ V_C(\x - { {\bf R}\over 2}) + V_C (\x + {{\bf R}\over 2}) \right]  
  e^{  i \x \cdot (\k-\q)  }   \nonumber \\
 & \times &   { \sqrt{ {(2j'_g + 1)(2j_g + 1)}} \over {16\pi} }
  \left[  D^{j'_g}_{\Lambda,\sigma'}(\hat\q) 
    D^{j_g,*}_{\Lambda\lambda'}(\hat\k) \chi^{\xi'}_{\sigma\sigma'} \chi^{\xi}_{\lambda\lambda'} 
    D^{1*}_{\mu\sigma}(\hat\q) D^1_{\mu\lambda}(\hat\k)  
     + \eta_Y\eta_Y' (\Lambda \to -\Lambda) \right] \left(\sqrt{ { \omega(k)} \over {\omega(q)}} 
   + \sqrt{ {\omega(q)} \over {\omega(k)}} \right)\nonumber \\
   & = & {N_C \over 2} 
\sum_{\lambda,\lambda',\sigma,\sigma',\mu} \int {{d\q} \over {(2\pi)^3}} \int {{d\k} \over {(2\pi)^3}} 
 \phi^*_{m',j'_g}(q) V_C(\k - \q)  \left[ e^{ - i {{\bf R} \over 2} \cdot (\k-\q)  } 
  +  e^{  i {{\bf R} \over 2} \cdot (\k-\q)  }  \right] \phi_{m,j_g}(k)
   \nonumber \\
 & \times &  
 {\sqrt{(2j'_g + 1)(2j_g + 1)} \over {16\pi} }   \left[  D^{j'_g}_{\Lambda,\sigma'}(\hat\q) 
    D^{j_g,*}_{\Lambda\lambda'}(\hat\k) \chi^{\xi'}_{\sigma\sigma'} \chi^{\xi}_{\lambda\lambda'} 
    D^{1*}_{\mu\sigma}(\hat\q) D^1_{\mu\lambda}(\hat\k)  
     + \eta_Y\eta_Y' (\Lambda \to -\Lambda) \right]  \left(\sqrt{ { \omega(k)} \over {\omega(q)}} 
   + \sqrt{ {\omega(q)} \over {\omega(k)}} \right), \nonumber \\
\end{eqnarray}
\end{widetext}
and $\eta_Y$ and $\xi$ related to $j_g$ and $\Lambda^Y_{PC}$  through  Eq.~(\ref{PC}). 

\begin{eqnarray}
H_{3d} & = &  -  {1\over {2N_C}} V_C(R)\delta_{m',m}\delta_{j'_g,j_g}  \nonumber \\
 &  = &  - 4\pi  {1\over {2N_C}}   \int {{dk k^2} \over {(2\pi)^3}} V_C(k) j_0(R k)\delta_{m',m}\delta_{j'_g,j_g} 
 \nonumber \\
 \end{eqnarray} 

\begin{widetext}
\begin{eqnarray}
H_{3e}  & = &   \sum \int {{d\k} \over {(2\pi)^3}} {{d\p} \over {(2\pi)^3}} {{d\q} \over {(2\pi)^3}} 
 {{\phi^*_{m',j'_g}(p)}\over {\sqrt{2\omega(p)}}} {{ \phi_{m,j_g}(k) }\over {\sqrt{2\omega(k)}}} 
  \nonumber \\ 
& \times &  \int d\x d\y d\z \left[  K(\x - {{\bf R}\over 2}, \z + \y - \x, \y + {{\bf R}\over 2})  + ({\bf R} \to - {\bf R}) \right] 
 e^{i\x \cdot \k} e^{i \z \cdot \q} e^{-i \y \cdot \p} \nonumber \\
 & \times &   { \sqrt{ {(2j'_g + 1)(2j_g + 1)}} \over {8\pi} } \left[
  D^{j'_g}_{\Lambda,\sigma'}(\hat\p) D^{1,*}_{\mu,\sigma}(\hat\p) \chi^{\xi'}_{\sigma'\sigma}
 D^1_{\mu,0}(\hat\q) 
  D^{j_g,*}_{\Lambda,\lambda'}(\hat\k) D^1_{\nu,\lambda}(\hat\k) \chi^{\xi}_{\lambda'\lambda}
 D^{1,*}_{\nu,0}(\hat\q)  + \eta_Y\eta'_Y (\Lambda \to -\Lambda) \right] \nonumber \\
 & = & 
 \sum \int {{d\k} \over {(2\pi)^3}} {{d\p} \over {(2\pi)^3}} {{d\q} \over {(2\pi)^3}} 
{{ \phi^*_{m',j'_g}(p)}\over {\sqrt{2\omega(p)}}}  {{\phi_{m,j_g}(k) }\over {\sqrt{2\omega(k)}}} K(\k+\q,\q,\p+\q)  
\left[  e^{i {{\bf R}\over 2} \cdot ( \k + \p + 2\q)} + ({\bf R} \to -{\bf R}) \right] 
\nonumber \\
 & \times &   { \sqrt{ {(2j'_g + 1)(2j_g + 1)}} \over {8\pi} } \left[
  D^{j'_g}_{\Lambda,\sigma'}(\hat\p) D^{1,*}_{\mu,\sigma}(\hat\p) \chi^{\xi'}_{\sigma'\sigma}
 D^1_{\mu,0}(\hat\q) 
  D^{j_g,*}_{\Lambda,\lambda'}(\hat\k) D^1_{\nu,\lambda}(\hat\k) \chi^{\xi}_{\lambda'\lambda}
 D^{1,*}_{\nu,0}(\hat\q)  + \eta_Y\eta'_Y (\Lambda \to -\Lambda) \right] \nonumber \\
 \end{eqnarray}
 \end{widetext}
 where the sum is over $\mu,\nu,\lambda,\lambda',\sigma,\sigma'$ and the kernel is given by 
 \begin{equation}
K(\x,\z,\y) 
 =   \int {{d\k} \over {(2\pi)^3}} {{d\p} \over {(2\pi)^3}} {{d\q} \over {(2\pi)^3}}
  K(k,q,p)  e^{i\x\cdot \k} e^{i\y \cdot \p} e^{i\z\cdot \q}  
 \end{equation}
 and 
 \begin{eqnarray}
 K(k,q,p) &  = & q^2 {{N_C^2 }\over {4}}  {{d(k) d(p) d(q)} \over {k^2q^2  p^2}}
 \nonumber \\
 & \times &   \left[ d(k) f(k) + d(p) f(p) + d(q) f(q) \right] \label{kk} 
 \end{eqnarray}
 Finally, 
 \begin{widetext}
\begin{eqnarray}
H_{3f}  & = &   \sum \int {{d\k} \over {(2\pi)^3}} {{d\p} \over {(2\pi)^3}} {{d\q} \over {(2\pi)^3}} 
 {{\phi^*_{m',j'_g}(p)}\over {\sqrt{2\omega(p)}}} {{ \phi_{m,j_g}(k)}\over {\sqrt{2\omega(k)}}} 
 \nonumber \\
& \times &   \int d\x d\y d\z \left[  K(\x - {{\bf R}\over 2}, \z + \y - \x, \y - {{\bf R}\over 2})  + ({\bf R} \to - {\bf R}) \right] 
 e^{i\x \cdot \k} e^{i \z \cdot \q} e^{-i \y \cdot \p} \nonumber \\
 & \times &   { \sqrt{ {(2j'_g + 1)(2j_g + 1)}} \over {8\pi} } \left[
  D^{j'_g}_{\Lambda,\sigma'}(\hat\p) D^{1,*}_{\mu,\sigma}(\hat\p) \chi^{\xi'}_{\sigma'\sigma}
 D^1_{\mu,0}(\hat\q) 
  D^{j_g,*}_{\Lambda,\lambda'}(\hat\k) D^1_{\nu,\lambda}(\hat\k) \chi^{\xi}_{\lambda'\lambda}
 D^{1,*}_{\nu,0}(\hat\q)  + \eta_Y\eta'_Y (\Lambda \to -\Lambda) \right] \nonumber \\
 & = & 
 \sum \int {{d\k} \over {(2\pi)^3}} {{d\p} \over {(2\pi)^3}} {{d\q} \over {(2\pi)^3}} 
{{ \phi^*_{m',j'_g}(p)}\over {\sqrt{2\omega(p)}}}{{ \phi_{m,j_g}(k)}\over {\sqrt{2\omega(k)}}}   K(\k+\q,\q,\p+\q)  
\left[  e^{i {{\bf R}\over 2} \cdot ( \k - \p )} + ({\bf R} \to -{\bf R}) \right] 
\nonumber \\
 & \times &   { \sqrt{ {(2j'_g + 1)(2j_g + 1)}} \over {8\pi} } \left[
  D^{j'_g}_{\Lambda,\sigma'}(\hat\p) D^{1,*}_{\mu,\sigma}(\hat\p) \chi^{\xi'}_{\sigma'\sigma}
 D^1_{\mu,0}(\hat\q) 
  D^{j_g,*}_{\Lambda,\lambda'}(\hat\k) D^1_{\nu,\lambda}(\hat\k) \chi^{\xi}_{\lambda'\lambda}
 D^{1,*}_{\nu,0}(\hat\q)  + \eta_Y\eta'_Y (\Lambda \to -\Lambda) \right] \nonumber \\
 \end{eqnarray}
 \end{widetext}
 In the large-$N_C$ limit, $g\sqrt{N_C} \sim O(1)$, and since $d(k) \propto g$ and $f  \sim O(1)$, all of  terms above are $O(1)$ except $H_d$ (which corresponds to a non-planar diagram, see  Fig.~\ref{vqqg}).  The  products of the three factors, $d(p_i)/p_i^2$,  originate from the three dressed Coulomb lines in diagrams $e$ and $f$ in 

Fig.~\ref{vqqg}, and the three  factors of  $f$ come from the three possibilities to insert the $\nabla^2$ operator on these three lines.  The derivative coupling  between transverse and Coulomb gluons leads to  the extra $q^2$ factor in the numerator in Eq.~(\ref{kk}). In coordinate space this  implies that $K(\x,\z,\y)$ is short-ranged in $\z$. Furthermore  in each of the three terms in Eq.~(\ref{kk}) there is only one combination, $d^2(p_i)f(p_i)/p_i^2$,  which  in momentum, space leads to the confining potential $V_C$. The remaining two are of the form  $d(p_i)/p_i^2$ with  $d(p) \propto 1/\sqrt{p}$, which for small momenta also leads to a short-ranged interaction decreasing as  $~1/\sqrt{r}$ for large $r$.  We thus conclude that for the three interaction lines connecting the four vertices in the "three-body force"  of Fig.~\ref{vqqg}e only one is long-ranged and all others are short-ranged.  Along these lines one can approximate $K(\x,\z,\y)$ as 

    \begin{equation}
    K(\x,\z,\y)  \propto \delta(\z)  \left[ {{m_g V_C(\x)} \over { (m_g |\y|)^\alpha}}  
     +  {{m_g V_C(\y)} \over {(m_g|\x|)^\alpha}}   \right],
    \end{equation}
 with $0<\alpha < 1$.   Ignoring the gluon spin and all spin-orbit couplings we then obtain,


\begin{eqnarray}
& & H_{3e}   \to  \int d\x d\y 
  {{\phi^*_{m'}(\x)}\over {\sqrt{2m_g}}} {{ \phi_{m}(\y)}\over {\sqrt{2m_g}}}  \nonumber \\
  & \times &  \left[  K(\x - {{\bf R}\over 2}, \y - \x, \y + {{\bf R}\over 2})  + ({\bf R} \to - {\bf R}) \right] 
  \nonumber \\ 
  & \propto &  \int d\x    \phi^*_{m'}(\x) \left[ { {V_C(\x - {{\bf R}\over 2}) } \over {(m_g|\x + {{\bf R}\over 2}|)^\alpha} } + ({\bf R} \to -{\bf R}) \right]  \phi_m(\x). \nonumber \\
  \end{eqnarray}

At large separation $R$ with the wave functions peaking at $|\x| \sim 0$, we find that $H_e$ grows less rapidly than  two-body interactions. This is in general true for interactions originating from the expansion of $K[A]$ in powers of $A$ which couple multiple gluons.  This is the basis for the  approximations discussed in Section.~\ref{chainsec}.

 The off-diagonal matrix element of the Hamiltonian mixing the $|\qq\rangle$ and $\qq g\rangle$ states, shown in Fig.~\ref{vmix}, is given by,
 
 \begin{widetext}
\begin{eqnarray}
H_{4}  & = &  i \sum \int {{d\k} \over {(2\pi)^3}}  {{d\q} \over {(2\pi)^3}} 
 {{\phi_{m,j_g}(k)}\over {\sqrt{2\omega(k)}}}   \int d\x d\z \left[  K_1(\x - {{\bf R}\over 2}, \z - \x - {{\bf R}\over 2})  - ({\bf R} \to - {\bf R}) \right] 
 e^{i\x \cdot \k} e^{i \z \cdot \q}  \nonumber \\
 & \times &   { \sqrt{ {2j_g + 1}} \over {4\pi} }
   D^{j_g,*}_{\Lambda=0,\lambda'}(\hat\k) D^1_{\nu,\lambda}(\hat\k) \chi^{\xi}_{\lambda'\lambda}
 D^{1,*}_{\nu,0}(\hat\q)  
  = 
 i \sum \int {{d\k} \over {(2\pi)^3}}  {{d\q} \over {(2\pi)^3}} 
 {{\phi_{m,j_g}(k)}\over {\sqrt{2\omega(k)}}}   K_1(\k+\q,\q)  \nonumber \\ 
& \times & \left[  e^{i {{\bf R}\over 2} \cdot ( \k + 2\q )} - ({\bf R} \to -{\bf R}) \right] 
 { \sqrt{ {2j_g + 1}} \over {4\pi} }
  D^{j_g,*}_{\Lambda=0,\lambda'}(\hat\k) D^1_{\nu,\lambda}(\hat\k) \chi^{\xi}_{\lambda'\lambda}
 D^{1,*}_{\nu,0}(\hat\q)   \nonumber \\
 \label{hvmix}
 \end{eqnarray}
 \end{widetext}
    
    \begin{equation}
    K_1(\x,\y) = \int {{d\p} \over {(2\pi)^3}} {{d\q} \over {(2\pi)^3}} K(p,q) 
 e^{i\x\cdot \p} e^{i\y \cdot \q}
    \end{equation}
    with 
 \begin{equation}
 K_1(p,q) = { {N_C\sqrt{C_F}}\over 2} q {{d(p)d(q)} \over {p^2q^2}}\left[d(p)f(p) + d(q)f(q)\right].
 \end{equation}

As expected in the large $N_C$ limit $K_1 = O(1)$ and  just like the three-body kernel described previously, $K_1(\x,\y)$ has mixed behavior for large separations.  A term, in momentum space, proportional to $d^2f$ in one of the two momentum variables leads to $V_C$ in the corresponding position space argument. While for the other momentum variable it leads to a less singular behavior for large distances. Approximately, we find 

\begin{equation} 
K_1(\x-{\R \over 2},\x+{\R \over 2}) \propto {{m_g^2 V_C(|\x - {{\bf R}\over 2}|) } \over {(m_g |\x + {{\bf R}\over 2}|)^\beta}} + ({\bf R} \to -{\bf R}) 
\end{equation}
with $1<\beta<2$. 
In this limit,  ignoring spin dependence, one finds 
\begin{equation}
H_{4} \to i \int d\x K_1(|\x - {{\bf R}\over 2}, |\x + {{\bf R}\over 2}|)  {{\phi_{m,j_g}(x)}\over {\sqrt{2 m_g
}}}. 
\end{equation}
Thus, similar to the case of $H_{3e}$, we find that at large separations  the mixing terms grow less rapidly with $R$ as compared to two-body interactions.


\begin{thebibliography}{99}

\bibitem{Juge:1997nc}
  K.~J.~Juge, J.~Kuti and C.~J.~Morningstar,
  Nucl.\ Phys.\ Proc.\ Suppl.\  {\bf 63}, 326 (1998)
  [arXiv:hep-lat/9709131].


\bibitem{Juge:2002br}
  K.~J.~Juge, J.~Kuti and C.~Morningstar,
  Phys.\ Rev.\ Lett.\  {\bf 90}, 161601 (2003)
  [arXiv:hep-lat/0207004].
  
\bibitem{Takahashi:2004rw}
  T.~T.~Takahashi and H.~Suganuma,
  Phys.\ Rev.\ D {\bf 70}, 074506 (2004)
  [arXiv:hep-lat/0409105].
  
\bibitem{Luscher:2004ib}
  M.~Luscher and P.~Weisz,
  JHEP {\bf 0407}, 014 (2004)
  [arXiv:hep-th/0406205].
  
\bibitem{Cornwall:2004gi}
  J.~M.~Cornwall,
  Phys.\ Rev.\ D {\bf 71}, 056002 (2005)
  [arXiv:hep-ph/0412201].
  
\bibitem{Greensite:2001nx}
  J.~Greensite and C.~B.~Thorn,
  JHEP {\bf 0202} (2002) 014
  [arXiv:hep-ph/0112326].
  
  
\bibitem{Bali:2005fu}
  G.~S.~Bali, H.~Neff, T.~Duessel, T.~Lippert and K.~Schilling  [SESAM
                  Collaboration],
  Phys.\ Rev.\ D {\bf 71}, 114513 (2005)
  [arXiv:hep-lat/0505012].
  
\bibitem{Bali:2000un}
  G.~S.~Bali,
  Phys.\ Rev.\ D {\bf 62}, 114503 (2000)
  [arXiv:hep-lat/0006022].
  
\bibitem{Juge:2004xr}
  K.~J.~Juge, J.~Kuti and C.~Morningstar,
  arXiv:hep-lat/0401032.
  
  
\bibitem{Greensite:2003bk}
  J.~Greensite,
  Prog.\ Part.\ Nucl.\ Phys.\  {\bf 51}, 1 (2003)
  [arXiv:hep-lat/0301023].


\bibitem{Juge:1999ie}
  K.~J.~Juge, J.~Kuti and C.~J.~Morningstar,
  Phys.\ Rev.\ Lett.\  {\bf 82}, 4400 (1999)
  [arXiv:hep-ph/9902336].

\bibitem{Juge:1999aw}
  K.~J.~Juge, J.~Kuti and C.~J.~Morningstar,
  Nucl.\ Phys.\ Proc.\ Suppl.\  {\bf 83}, 304 (2000)
  [arXiv:hep-lat/9909165].
  
  
\bibitem{Hasenfratz:1980jv}
  P.~Hasenfratz, R.~R.~Horgan, J.~Kuti and J.~M.~Richard,
  Phys.\ Lett.\ B {\bf 95}, 299 (1980).
  
\bibitem{Juge:1997nd}
  K.~J.~Juge, J.~Kuti and C.~J.~Morningstar,
  Nucl.\ Phys.\ Proc.\ Suppl.\  {\bf 63}, 543 (1998)
  [arXiv:hep-lat/9709132].
  
\bibitem{Isgur:1984bm}
  N.~Isgur and J.~Paton,
  Phys.\ Rev.\ D {\bf 31}, 2910 (1985).
  
\bibitem{Horn:1977rq}
  D.~Horn and J.~Mandula,
  Phys.\ Rev.\ D {\bf 17}, 898 (1978).
  
\bibitem{Swanson:1998kx}
  E.~S.~Swanson and A.~P.~Szczepaniak,
  Phys.\ Rev.\ D {\bf 59}, 014035 (1999)
  [arXiv:hep-ph/9804219].
  
\bibitem{Juge:2003ge}
  K.~J.~Juge, J.~Kuti and C.~Morningstar,
  arXiv:hep-lat/0312019.
  
\bibitem{Thorn:1979gu}
  C.~B.~Thorn,
  Phys.\ Rev.\ D {\bf 20}, 1435 (1979).
  
\bibitem{Szczepaniak:1996tk}
  A.~P.~Szczepaniak and E.~S.~Swanson,
  Phys.\ Rev.\ D {\bf 55}, 3987 (1997)
  [arXiv:hep-ph/9611310].
  
\bibitem{Greensite:2003xf}
  J.~Greensite and S.~Olejnik,
  Phys.\ Rev.\ D {\bf 67}, 094503 (2003)
  [arXiv:hep-lat/0302018].
  
\bibitem{Greensite:2004ke}
  J.~Greensite, S.~Olejnik and D.~Zwanziger,
  Phys.\ Rev.\ D {\bf 69}, 074506 (2004)
  [arXiv:hep-lat/0401003].
  
  
\bibitem{Zwanziger:2002sh}
  D.~Zwanziger,
  Phys.\ Rev.\ Lett.\  {\bf 90}, 102001 (2003)
  [arXiv:hep-lat/0209105].
  
\bibitem{Christ:1980ku}
  N.~H.~Christ and T.~D.~Lee,
  Phys.\ Rev.\ D {\bf 22}, 939 (1980)
  [Phys.\ Scripta {\bf 23}, 970 (1981)].
  
\bibitem{vanBaal:1997gu}
  P.~van Baal,
  arXiv:hep-th/9711070.
  
\bibitem{Reinhardt:2004mm}
  H.~Reinhardt and C.~Feuchter,
  Phys.\ Rev.\ D {\bf 71}, 105002 (2005)
  [arXiv:hep-th/0408237].
  
\bibitem{Feuchter:2004mk}
  C.~Feuchter and H.~Reinhardt,
  Phys.\ Rev.\ D {\bf 70}, 105021 (2004)
  [arXiv:hep-th/0408236].

  
  
\bibitem{Szczepaniak:2003ve}
  A.~P.~Szczepaniak,
  Phys.\ Rev.\ D {\bf 69}, 074031 (2004)
  [arXiv:hep-ph/0306030].
  
  
\bibitem{Szczepaniak:2001rg}
  A.~P.~Szczepaniak and E.~S.~Swanson,
  Phys.\ Rev.\ D {\bf 65}, 025012 (2002)
  [arXiv:hep-ph/0107078].
  
\bibitem{Langfeld:2004qs}
  K.~Langfeld and L.~Moyaerts,
  Phys.\ Rev.\ D {\bf 70}, 074507 (2004)
  [arXiv:hep-lat/0406024].
  
  
  
\bibitem{Adler:1984ri}
  S.~L.~Adler and A.~C.~Davis,
  Nucl.\ Phys.\ B {\bf 244}, 469 (1984).
  
\bibitem{Szczepaniak:1995cw}
  A.~Szczepaniak, E.~S.~Swanson, C.~R.~Ji and S.~R.~Cotanch,
  Phys.\ Rev.\ Lett.\  {\bf 76}, 2011 (1996)
  [arXiv:hep-ph/9511422].
  
\bibitem{Ligterink:2003hd}
  N.~Ligterink and E.~S.~Swanson,
  Phys.\ Rev.\ C {\bf 69}, 025204 (2004)
  [arXiv:hep-ph/0310070].
  
\bibitem{Szczepaniak:2003mr}
  A.~P.~Szczepaniak and E.~S.~Swanson,
  Phys.\ Lett.\ B {\bf 577}, 61 (2003)
  [arXiv:hep-ph/0308268].
  

  \bibitem{colin} C.~J.~Morningstar, {\it private communication} 


\end{thebibliography}
\end{document}